\documentclass[aps,prl,floatfix,twocolumn,showpacs]{revtex4}
\usepackage{epsfig}
\usepackage{verbatim}	 
\usepackage{amssymb}
\usepackage{bbold}
\usepackage{amsmath}
\usepackage{psfrag}
\usepackage{color}
\usepackage{ulem}
\newcommand{\bi}{\bibitem}
\usepackage[latin1]{inputenc}
\usepackage[english]{babel}
\usepackage{graphicx}
\usepackage{amscd}
\usepackage{eucal}
\usepackage{bm}
\input {epsf}
\begin{document}

\title{Signatures of Majorana fermions in hybrid normal--superconducting rings}
\author{Ph.~Jacquod$^{1,2}$ and M.~B\"uttiker$^3$}

\affiliation{$^1$Physics Department, University of Arizona,
Tucson, AZ 85721, USA \\
$^2$ College of Optical Sciences, University of Arizona, Tucson, AZ 85721\\
$^3$Theoretical Physics Department, University of Geneva, 1211 Geneva, Switzerland}

\date{26th June 2013}

\begin{abstract}
We investigate persistent currents in metallic rings interrupted by a Coulomb blockaded topological superconducting segment.
We show that the presence of Majorana bound states in the superconductor is reflected in the emergence of an
$h/e$ harmonics in the persistent current, whose sign is determined by the fermion parity of the superconductor.
The Majorana bound states further render the current finite at zero flux, nevertheless 
the resulting peculiar symmetry of the persistent current is compatible with a free energy that is even in time-reversal 
symmetry breaking fields. These unique features of the persistent currents are robust against disorder
and provide unambiguous signatures of the presence of Majorana fermions.  
\end{abstract}

\maketitle

Several recent experiments
have reported features in transport~\cite{Mou12,Den12,Das12,Fin12}
and Josephson spectroscopy~\cite{Rok12} that have varying degrees of consistency with the presence of 
Majorana states~\cite{Ali12}. The experimental setups are all based on the
nanowire implementation of Kitaev's chain~\cite{Kit01} proposed in Refs.~\cite{Lut10,Ore10}.
The presence of Majorana fermions in such systems manifests itself in a zero-bias peak in the tunneling conductance
into the nanowire~\cite{Law09,Akh11} and by a doubling of the periodicity of the Josephson current in the 
superconducting phase difference~\cite{Kit01,Fu09a,Pie13}. These features were observed in 
Refs.~\cite{Mou12,Den12,Das12,Fin12,Rok12}. Other 
signatures of Majorana states in transport interferometry have also been theoretically 
investigated~\cite{Akh11,Fu09b,Ben10,Jia12}.

Motivated by these experimental reports, a number of theoretical works have pointed out that 
zero-bias peaks in the tunneling conductance may also occur in the topologically trivial phase~\cite{Bag12,Liu12,Pik12}.
Their observation is therefore not sufficient to demonstrate the presence of a Majorana state.
Moreover, Ref.~\cite{Sau12} showed that a doubling of the periodicity of the Josephson current
as reported in the AC Josephson setup of Ref.~\cite{Rok12}
may also occur due to Landau-Zener transitions between standard Andreev bound states. Despite 
a slowly growing body of evidence in favor of the presence of Majorana states in nanowires, a
smoking gun experiment is still missing. The consensus at this point is that only 
a zero-bias tunneling conductance peak quantized at $2e^2/h$ would unambiguously reflect the 
presence of a Majorana state. The observation of such a quantized peak 
would however require ideal circumstances, in particular
ultra-low temperatures beyond the reach of currently existing devices~\cite{Fra13}. 

\begin{figure}
\psfrag{phi}{\LARGE $ \phi$}
\psfrag{xi1}{$\xi_1$}
\psfrag{xi2}{$\xi_2$}
\psfrag{sc}{topological superconductor}
\psfrag{no}{normal metallic ring}
\psfrag{x}{$x$}
\psfrag{y}{$y$}
\psfrag{y1}{$y_1$}
\psfrag{y2}{$y_2$}
\psfrag{z}{$z$}
\psfrag{V}{$V_g$}
\psfrag{Bz}{\color{red} $B_{\rm z}$}
\centering
 \includegraphics[width=0.99\columnwidth]{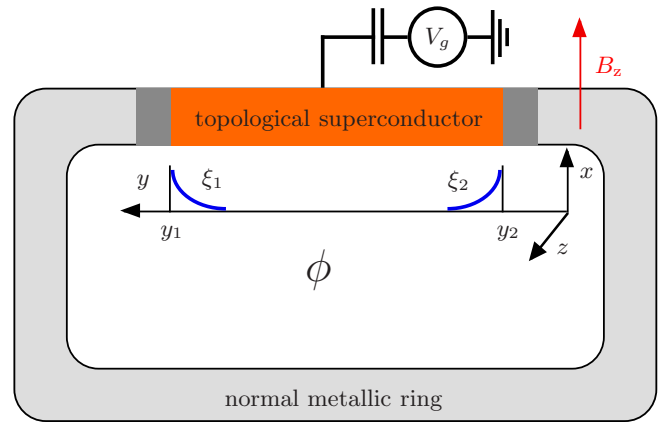}
\caption{\label{fig:fig1} (Color online) Setup to detect signatures of Majorana fermions in persistent currents through 
metallic rings.
A spin-orbit coupled metallic nanowire (orange) with induced superconductivity is embedded in a metallic ring pierced by
a magnetic flux $\phi$. A sufficiently strong Zeeman field $B_{\rm Z}$ applied parallel to the axis of the nanowire creates
two Majorana states, $\xi_{1,2}$, localized at each end of the nanowire. In the presence of such states, individual electrons can
be coherentely transferred across the nanowire, even when the latter is longer than the superconducting coherence length,
thereby generating a $h/e$ harmonics in the persistent current. The nanowire is Coulomb blockaded and tunnel-coupled to the 
metallic ring (dark grey rectangles represent tunnel barriers). Its occupation
number $n_0$ can be externally tuned by a gate voltage $V_g$, which fixes the fermion parity and
allows to change the sign of the persistent current
$I(\phi) \propto (-1)^{n_0}$.}
\end{figure}

In this manuscript, we propose an altogether new experiment to detect Majorana
fermions in the nanowire platforms of Refs.~\cite{Lut10,Ore10}. The system we investigate is sketched in Fig.~\ref{fig:fig1}.
It consists of a normal metallic ring interrupted by a superconducting segment of length $L \gg \xi$, much larger than 
the superconducting coherence length $\xi$. The superconducting segment can be either in a topologically
trivial state, with a superconducting gap which allows the transfer of Cooper pairs only, or in a topologically
nontrivial state, with Majorana subgap states at each of its ends allowing the coherent transfer of single 
quasiparticles~\cite{Nil08,Fu10}. A persistent current is triggered by a magnetic flux
$\phi$ piercing the ring. In the former case, the current has 
even harmonics only $\sim \sin(4 \pi n \phi/\phi_0)$, with the flux quantum $\phi_0=h/e$ and $n=1,2,...$
In other words, only $h/2e$ harmonics exist, because only Cooper pairs with charge $2e$ can be transferred through the superconductor. 
In the latter case, the transfer of a single electron generates odd harmonics $\sim \sin(2 \pi (2 n +1) \phi/\phi_0-\chi)$.
The presence of Majorana states breaks time-reversal symmetry, because the quasiparticle transfer amplitude $\tau$ through the 
topological superconductor picks a relative minus 
sign when the direction of transfer is inverted. This follows directly from the fermionic anticommutation relations of Majorana 
creation and annihilation operators~\cite{Nil08,Fu10}. Additionally, 
$\tau \sim (-1)^{n_0}$ depends on the number $n_0$ of fermions on the superconductor. 
This leads to $\chi=(-1)^{n_0} \pi/2$ and, when the fermion parity is fixed, persistent current harmonics
$\sim (-1)^{n_0} \cos(2 \pi (2 n +1) \phi/\phi_0)$ are obtained in the weak coupling limit, while the 
odd harmonics vanish identically when the fermion parity is not fixed and an average over $n_0$ is taken. 
Building up on Ref.~\cite{Fu10}, we propose to fix the fermion parity via Coulomb blockade of the superconducting segment. 
The on-resonance persistent current then bears three unambiguous signatures of the presence of Majorana states:
(i) the current develops a $h/e$ harmonic,
(ii) the current is finite at zero flux in the nontrivial phase but vanishes in the trivial phase, and
(iii) the current changes sign when the fermion number parity on the superconductor is
changed from $n_0=2 n$ (unoccupied Majorana states) to $n_0=2n +1$ (occupied Majorana states).
We finally show that despite
the finite zero-field current, the system's free energy is even in time-reversal symmetry breaking fields,
$\mathcal{F}(\phi,B_{\rm Z}) = \mathcal{F}(-\phi,-B_{\rm Z})$. 

The origin of the $h/e$ harmonics is the same as that of the fractional Josephson effect, however 
the effect is more robust for persistent currents than for the AC Josephson
effect~\cite{Rok12}, the former in particular are immune to Landau-Zener transitions~\cite{Sau12}. 
Measuring persistent currents in metallic rings is challenging but has been done by several groups~\cite{Lev90,Cha91,Mai93,Blu09,Ble09}.
In persistent currents the emergence of a $h/e$ harmonic may also occur when $\xi$ increases and becomes comparable
to $L$~\cite{But86}, which however can be monitored experimentally. 
Anomalous supercurrents (a non-zero persistent current 
at zero phase difference) in Josephson junctions formed of a mesoscopic conductor 
sandwiched between two ordinary superconductors are discussed in several theoretical works (See e.g. Refs.~\cite{Rey08,Zaz09,Bru13}). 
The anomalous supercurrent is a consequence of the spin-orbit interaction and Zeeman fields in the normal conductor. 
In contrast, in our work spin-orbit interaction and Zeeman fields exist only in the superconductor to the extent that they 
are needed to drive the latter into the topological phase. 


We calculate the canonical persistent current using the effective Hamiltonian derived by Fu~\cite{Fu10}
for fixed-parity, Coulomb-blockaded topological superconductors,
\begin{eqnarray}\label{eq:effH}
H &=& H_{\rm ring} + \delta (f^\dagger f-1/2) + 
[\lambda_1 c_{\rm L}^\dagger f + h.c.] \, \nonumber \\
&+& 
[-i \lambda_2  \, (-1)^{f^\dagger f} 
\, c_{\rm R}^\dagger f \exp[i \varphi] + h.c.]  \, .
\end{eqnarray}
The superconductor is connected to a metallic ring with Hamiltonian $H_{\rm ring} = \sum_{k} \epsilon_k c^\dagger_k c_k$, 
pierced by  a magnetic flux $\phi =\hbar \varphi/e$, the fermionic operators
$c$ annihilate an electron in the ring, $f$ is a fermion operator on the superconductor,
combining Majorana operators and fermion parity isospin operators. The superconductor is Coulomb blockaded and coupled
to the ring via tunnel barrier, $\lambda_{1,2} \ll t$, which
allows to project the Hamiltonian into the subspace with only two superconductor charge states 
$|n_0\rangle$ and $|n_0+1\rangle$. The energy difference between these two states is $\delta$, which can be tuned
by an external gate voltage. 
Details of the construction of $H$ are given in Ref.~\cite{Fu10}.
The above Hamiltonian is appropriate to calculate the $h/e$ harmonics of the persistent current close
to zero chemical potential in the ring.

In the limit when the superconductor-ring coupling is weak, and close to resonance between $|n_0\rangle$ and $|n_0+1\rangle$ (i.e. 
close to $\delta=0$) with $M+n_0$ electrons in total, $H$ can be reduced to a 2 $\times$ 2 Hamiltonian (see Ref.~\cite{But96}
for a similar treatment of a metallic quantum dot embedded in a metallic ring)
\begin{equation}\label{eq:hred}
H_{\rm red} = \left(
\begin{array}{cc}
\epsilon_{M} & \tilde{\lambda}_1  -  i  \tilde{\lambda}_2 (-1)^{n_0} e^{i \varphi}\\
 \tilde{\lambda}_1 +  i  \tilde{\lambda}_2 (-1)^{n_0} e^{- i \varphi} & \delta 
 \end{array}
 \right) \, .
\end{equation}
The total energy is given by the eigenvalues of $H_{\rm red}$ plus a constant sum over
all energy levels in the ring $\sum_{k=1}^{M-1} \epsilon_k$. The eigenvalues
$E_\pm(\varphi)$ of $H_{\rm red}$ are easily 
calculated, and one obtains the canonical persistent current
$I(\varphi) = -\partial_\phi E_-$ as
\begin{equation}\label{eq:pcurrent}
I(\varphi) = 
\frac{e}{\hbar}  \frac{(-1)^{n_0} \tilde{\lambda}_1 \tilde{\lambda}_2 \, \cos \varphi}{\sqrt{(\epsilon_{M}-\delta)^2/4
+ \tilde{\lambda}_1^2+\tilde{\lambda}_2^2+2 \tilde{\lambda}_1 \tilde{\lambda}_2 (-1)^{n_0} \sin\varphi}}\, .
\end{equation}
The $h/e$ harmonics of the current depends on the fermion parity, it is finite at zero flux and its magnitude is  
algebraically reduced away from resonance ($\delta \ne \epsilon_M$). The weak coupling condition means that
$\tilde{\lambda} \ll \delta \epsilon$ with the energy level spacing $\delta \epsilon$ in the ring, so that unless
$\epsilon_M$ is anomalously close to zero, 
the $\tilde{\lambda}$-terms under the square root in the denominator are negligible at the degeneracy point between the two superconducting
states, $\delta=0$. Generically, ${\rm min}(|\epsilon_M|) \simeq \delta \epsilon$. Further specifying to 
a weakly disordered, $N$-site, one-dimensional tight-binding ring with nearest neighbor hopping, 
we obtain $\tilde{\lambda}_i = (\pi/2 N)^{1/2} \lambda_i$, $\delta \epsilon \simeq \pi t/N$ close to $\epsilon_M=0$,
so that $I(\varphi)=(e/h) (-1)^{n_0} \lambda_1 \lambda_2 \, \cos \varphi /t$. In the weak coupling limit, the sign of the 
current is thus determined by the fermion parity, which can be tuned via the gate voltage. The change in sign is a signature
of the Friedel sum rule~\cite{Fri52}. In this weak coupling limit, the current
does not depend on the length of the ring. 
Alternatively, one may tune the gate voltage and work at $\epsilon_M = \delta$, in which case the current becomes
(for $\lambda_{1,2}=\lambda$)
$I(\varphi) = (e/ \hbar) (-1)^{n_0} \pi^{1/2} \lambda  \, \cos \varphi \big/ \sqrt{4 N[1+(-1)^{n_0} \sin\varphi]}$.
In this case, the current decays as $N^{-1/2}$ with the size of the ring. 
One important consequence of Eq.~(\ref{eq:pcurrent}) is that upon disorder averaging, the $h/e$ harmonics vanishes identically
because $I_{n_0}+I_{n_0+1}$ is $h/2e$ periodic.

There are two main differences between Eq.~(\ref{eq:pcurrent}) 
and Eq.~(7) of Ref.~\cite{But96} for a ring interrupted by a normal 
metallic quantum dot. The first one is that
the presence of Majorana states breaks time-reversal symmetry in such a way that $\varphi \rightarrow \varphi \pm \pi/2$, thereby 
turning sines into cosines and vice-versa. The second is
that in Eq.~(\ref{eq:pcurrent}), the parity of the current is related to the fermion parity
on the superconductor when the latter is in the nontrivial phase. 
In the case of a purely one-dimensional ring, the parity of the persistent current depends on the 
total parity of the one-dimensional ring, including both the Coulomb blockaded segment and the metallic ring~\cite{But96}. 
This remains true also in the case of a superconducting ring, because the occupancy of the metallic ring determines the
relative sign of the hoppings $\lambda_1$ and $\lambda_2$. Parity effects disappear in non strictly one-dimensional metallic rings, however 
the dependence on the fermion parity of the superconducting segment persists, even when the metallic ring carries
more than one transverse channel. 

\begin{figure}
\psfrag{I}{   $I(\varphi) t/\lambda^2 \;\; [e/\hbar]$}
\psfrag{f}{$\varphi$}
\centering
 \includegraphics[width=0.9\columnwidth]{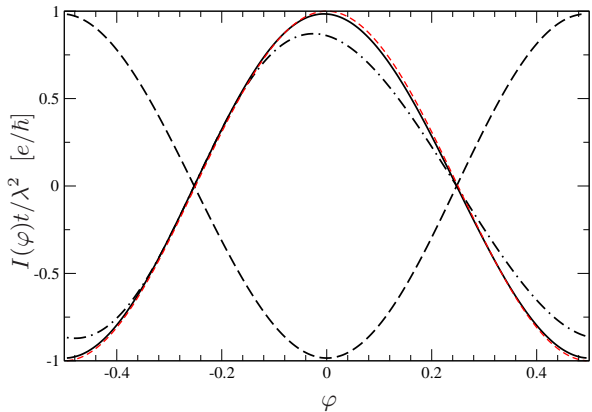}
\caption{\label{fig:fig2} (Color online) Persistent currents in the set-up of Fig.~\ref{fig:fig1} with weak coupling
$\lambda_1=\lambda_2=0.02 t$ between the topological superconductor and the metallic ring. The metallic ring has
$N=20$ (solid lines)  $100$ (dashed line) $500$ (dotted line) and $1500$ (dotted-dashed line) sites. The fermion parity $n_0$ is even
(black lines) and odd (blue line). The red dashed line gives the theoretical prediction
$I(\varphi) = (e/\hbar) (-1)^{n_0} \lambda_1 \lambda_2 \cos \varphi$ (see text). Deviations appear for larger
rings with smaller level spacing for which $\tilde{\lambda_i}$ is no longer negligible against $\delta \epsilon$.}
\end{figure}

To check these results and extend them beyond the weak coupling limit we numerically calculate $I(\varphi)$
for a tight-binding version of Eq.~(\ref{eq:effH}) with
$H_{\rm ring} = -t \sum_{\langle i;j\rangle} (c_i^\dagger c_j + h.c.) $ and $c_{\rm L} = c_1$, $c_{\rm R} = c_N$. 
We first show in Fig.~\ref{fig:fig2} results for the weak coupling regime with $\lambda_i \ll \delta \epsilon$. As predicted,
the persistent current exhibits a simple $\cos \varphi$-behavior, with a sign determined by the fermion parity as
$(-1)^{n_0}$. Upon increasing the size of the ring, the level spacing $\delta \epsilon$ decreases and therefore
the $\lambda$-terms  in the denominator of Eq.~(\ref{eq:pcurrent}) are no longer entirely
negligible against the $(\epsilon_M-\delta)^2$ term. This generates
deviations from the pure $\cos \varphi$ behavior (dashed, dotted and dotted-dashed curve). 
Note that for clean one-dimensional rings with odd number of sites, 
one eigenenergy vanishes, in which case the $\lambda$-terms are not negligible at resonance, $\delta=0$.
We checked, but do not show, that this anomalous even-odd effect  generically disappears when $\delta$
is tuned slightly away from resonance, when some disorder is added to the ring or when the latter becomes
quasi-one-dimensional, with few transverse channels. 

\begin{figure}
\psfrag{I}{   $I(\varphi)/\lambda \;\; [e/\hbar]$}
\psfrag{f}{$\varphi$}
\centering
 \includegraphics[width=0.9\columnwidth]{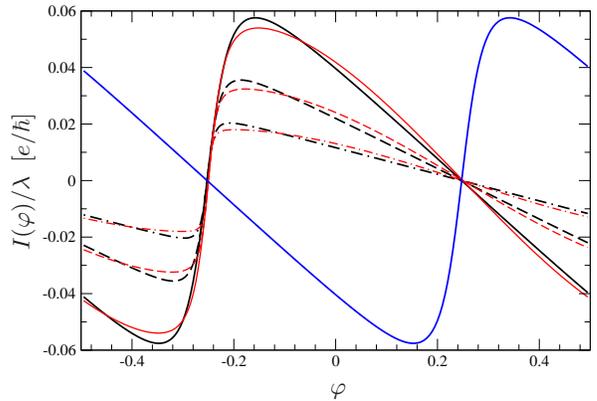}
\caption{\label{fig:fig3} (Color online)  Black and blue lines: persistent currents in the set-up of Fig.~\ref{fig:fig1} with larger 
coupling $\lambda_1=\lambda_2=0.2 t$ between the topological superconductor and the metallic ring.
The metallic ring has $N=100$ (solid line), 200 (dashed line) and 400 (dotted-dashed line) sites, with even 
(black lines) and odd (blue line, for $N=100$) fermion parity. The red lines are best fits with Eq.~(\ref{eq:pcurrent}),
which deviate slightly from the theoretical prediction because more than a single level is coupled
to the Majorana states for the chosen set of parameters in the numerical simulations.}
\end{figure}

Data at stronger couplings $\lambda_1=\lambda_2=0.2 t$ are shown in Fig.~\ref{fig:fig3}. The persistent currents acquires
a more complicated harmonics content, which is qualitatively captured by Eq.~(\ref{eq:pcurrent}). Some deviations
are seen, which we attribute to the fact that more than a single level is coupled to the superconductor in this regime,
while Eq.~(\ref{eq:pcurrent}) assumes a single-level hybridization. For the same reason, the best fits with 
Eq.~(\ref{eq:pcurrent}) shown have parameters that differ from the theoretical predictions. It is still remarkable
that the single-level prediction of Eq.~(\ref{eq:pcurrent}) captures the current symmetry
$I(n_0,\varphi)=-I(n_0+1,-\varphi)$, even when more than one ring level is coupled to the Majorana states. 

Our numerics confirm our theoretical prediction that in clean, one-dimensional metallic rings interrupted
by a Coulomb-blockaded superconducting segment carrying Majorana bound states, persistent currents have
$h/e$ periodicity and finite magnitude even when no flux pierces the ring, and furthermore change
parity with the fermion number parity on the superconductor. These predictions still hold, even when the 
metallic ring is disordered and carries more than a single channel. This can be seen in Eq.~(\ref{eq:pcurrent}) which remains 
valid regardless of the microscopic Hamiltonian giving the eigenvalue $\epsilon_M$.
Impurities can also exist inside the superconductor, in which case they can
generate nontopological domains with Majorana bounds states at the domain walls~\cite{Shi10,Fle10}. This can be modeled
by replacing the single Majorana site in the effective Hamiltonian of Eq.~(\ref{eq:effH}) with a chain of $p$ Majorana
sites connected by disordered normal-metallic segments. It is then easy to show using a gauge transformation on the 
fermionic operators $c$ and $f$ that the 
$h/e$ harmonics of the persistent current becomes $\sim \sin(2 \pi (2 n +1) \phi/\phi_0-\chi)$
with $\chi = (-1)^{n_{\rm tot}} p \pi/2$. Therefore, the general form of Eq.~(\ref{eq:pcurrent}) is preserved when there are an odd number of
Majorana-carrying segments in the chain, while $\cos \leftrightarrow \sin$ when $p$ is even. Either way, the $h/e$ harmonics
persists and changes parity with the parity of the sum of the number of fermions on all Majorana segments.

The persistent current is given by $I(\phi) = -\partial_\phi \mathcal{F}$ with the free energy $\mathcal{F}$. 
In the absence of magnetization, the latter must be an even function of magnetic field, therefore one would 
expect $I(\phi=0)=0$. The finite zero-flux current predicted above comes about because creating
a Majorana state requires to break time-reversal symmetry with a Zeeman magnetic field in the first place,
and the free energy is even in the total field, $\mathcal{F}(\phi,B_{\rm Z}) = \mathcal{F}(-\phi,-B_{\rm Z})$.
To show that this is the case, we
specify to the standard nanowire Hamiltonian for Majorana bound states~\cite{Lut10,Ore10}, 
\begin{eqnarray}
H &=& [p_y^2/2m - \mu] \tau_z + u p_y \sigma_z \tau_z + B_{\rm Z} \sigma_x + \Delta \tau_x \, .
\end{eqnarray}
The wire is aligned in the $y$-direction, $\sigma_\alpha$ and $\tau_\alpha$
are Pauli matrices in spin and particle-hole space, respectively. Inverting $B_{\rm Z}$
is equivalent to space inversion
$y \rightarrow -y$, $p_y \rightarrow -p_y$, $\sigma_y \rightarrow \sigma_y$,
$\sigma_{x,z} \rightarrow -\sigma_{x,z}$ and $\tau_\alpha \rightarrow \tau_\alpha$. Therefore, 
$B_{\rm Z} \rightarrow -B_{\rm Z}$ is equivalent to interchanging the Majorana operators. 
This is equivalent to the substitution $\lambda_1 c_1 \leftrightarrow \lambda_2 c_N$ in our tight-binding formulation of 
Eq.~(\ref{eq:effH}), which can 
be absorbed by a relabelling the ring operators $c_i \rightarrow c_{N-i+1}$ together with 
$\phi \rightarrow -\phi$ (because the relabelling inverts the direction of counting sites along the ring).
Thus the free energy is even, $F(\phi,B_{\rm Z}) = F(-\phi,-B_{\rm Z})$ 
and the persistent current is odd in the total magnetic field, as should be.

It is interesting to note that these symmetry considerations translate into an apparent violation of the 
Onsager reciprocity relation $G(\phi) \ne G(-\phi)$ for the conductance of the system when it is connected
to two external leads. Such an apparent violation has been reported in Ref.~\cite{Ben10}, but its origin was
not discussed. The above argument for the symmetry of the free energy can be applied to the transport
problem, giving the true Onsager symmetry $G(\phi,B_{\rm Z}) = G(-\phi,-B_{\rm Z})$, 
containing both the Aharonov-Bohm flux and the time-reversal symmetry breaking field generating the 
Majorana states in the first place. Such reciprocity relations cannot be violated in two-terminal systems in the linear
response regime. Simultaneously, the oscillating part of $G(\phi)$ changes sign each time an electron is added
on the superconductor, which agrees with the Friedel sum rule~\cite{Fri52}, according to which the scattering phase 
jumps by $\pi$ each time the energy of the
scattering particle crosses a quasi-bound state of the scatterer. Topological superconductors therefore
present a unique opportunity to verify this sum rule in the presence of superconductivity. In the topologically 
trivial regime, only Cooper pairs can be added, which result in unnoticeable scattering phase jumps of $2 \pi$. 

We thank C. Beenakker, B. Sothmann and M. Wimmer for interesting discussions. PJ thanks the Lorentz Institute,
Leiden University and the Theoretical Physics Department, University of Geneva
for their hospitality at the early stage of this project. Research in Geneva was supported by the 
Swiss NSF, the NCCR MaNEP and QSIT.


\begin{thebibliography}{9}

\bi{Mou12} V. Mourik, K. Zuo, S.M. Frolov, S.R. Plissard, E.P.A.M. Bakkers, and L.P. Kouwenhoven, 
Science {\bf 336}, 1003 (2012).
 
\bi{Den12} M.T. Deng, C.L. Yu, G.Y. Huang, M. Larsson, P. Caroff,
and H.Q. Xu, Nano Lett. {\bf 12}, 6414 (2012).

\bi{Das12} A. Das, Y. Ronen, Y. Most, Y. Oreg, M. Heiblum, and H. Shtrikman, Nature Phys. {\bf 8}, 887 (2012).

\bi{Fin12} A.D.K. Finck, D.J. Van Harlingen, P.K. Mohseni, K. Jung, and X. Li, Phys. Rev. Lett. {\bf 110}, 
126406 (2013).

\bi{Rok12} L.P. Rokhinson, X. Lui, and J.K. Furdyna, Nature Phys. {\bf 8}, 795 (2012).

\bi{Ali12} For reviews on Majorana fermions in a condensed matter context, see: J. Alicea, 
Rep. Prog. Phys. {\bf 75}, 076501 (2012); C.W.J. Beenakker, Annu. Rev. Con. Mat. Phys. {\bf 4}, 113 (2013).

\bi{Kit01} A.Y. Kitaev, Phys. Usp. {\bf 44}, 131 (2001).

\bi{Lut10} R.M. Lutchyn, J.D. Sau, and S. Das Sarma, Phys. Rev. Lett. {\bf 105}, 077001 (2010).

\bi{Ore10} Y. Oreg, G. Refael, and F. von Oppen, Phys. Rev. Lett. {\bf 105}, 177002 (2010).

\bi{Law09}  K.T. Law, P.A. Lee, and T.K. Ng, Phys. Rev. Lett. {\bf 103}, 237001 (2009). 
 
\bi{Akh11}  A. R. Akhmerov, J. P. Dahlhaus, F. Hassler, M. Wimmer, and C. W. J. Beenakker
Phys. Rev. Lett. 106, 057001 (2011).

\bi{Fu09a} L. Fu and C.L. Kane, Phys. Rev. B {\bf 79}, 161408 (2009).

\bi{Pie13} F. Pietka, A. Romito, M. Duckheim, Y. Oreg, and F. von Oppen, New J. Phys. {\bf 15}, 025001 (2013).

\bi{Fu09b} L. Fu and C.L. Kane, Phys. Rev. lett. {\bf 102}, 216403 (2009).

\bibitem{Ben10} C. Benjamin and J.K. Pachos, Phys. Rev. B {\bf 81}, 085101 (2010).

\bi{Jia12} J. Li, G. Fleury, and M. B\"uttiker, Phys. Rev. B {\bf 85}, 125440 (2012).

\bi{Bag12} D. Bagrets and A. Altland, Phys. Rev. Lett. {\bf 109}, 227005 (2012).

\bi{Liu12} J. Liu, A.C. Potter, K.T. Law, and P.A. Lee, Phys. Rev. Lett. {\bf 109}, 267002 (2012).

\bi{Pik12} D.I. Pikulin, J.P. Dalhaus, M. Wimmer, H. Schomerus, and C.W.J. Beenakker, 
New J. Phys. {\bf 14}, 125011 (2012).

\bi{Sau12} J.D. Sau, E. Berg, and B.I. Halperin, arXiv:1206.4596.

\bi{Fra13} M. Franz,  Nature Nanotechnology {\bf 8}, 149 (2013).

\bibitem{Nil08} J. Nilsson, A.R. Akhmerov, and C.W.J. Beenakker, Phys. Rev. Lett. {\bf 101},
120403 (2008).

\bibitem{Fu10} L. Fu, Phys. Rev. Lett. {\bf 104}, 056402 (2010).

\bibitem{Lev90} L.P. L\'evy, G. Dolan, J. Dunsmuir, and H. Bouchiat, Phys. Rev. Lett. {\bf 64}, 2074 (1990).

\bi{Cha91} V. Chandrasekhar, R.A. Webb, M.J. Brady, M.B. Ketchen, W.J. Gallagher, and A. Kleinsasser,
Phys. Rev. Lett. {\bf 67}, 3578 (1991).

\bi{Mai93} D. Mailly, C. Chapelier, A. Benoit, Phys. Rev. Lett. {\bf 70}, 2020 (1993).

\bi{Blu09} H. Bluhm, N.C. Koschnick, J.A. Bert, M.E. Huber, and K.A. Moler, Phys. Rev. Lett. {\bf 102},
136802 (2009).

\bi{Ble09} A.C. Bleszynski-Jayich, W.E. Shanks, B. Peaudecerf, E. Ginossar, F. von Oppen, L. Glazman, and
J.G.E. Harris, Science {\bf 326}, 272 (2009).

\bibitem{But86} M. B\"uttiker and T. M. Klapwijk, Phys. Rev. B {\bf 33}, 5114 (1986).

\bi{Rey08} A.A. Reynoso, G. Usaj, C.A. Balseiro, D. Feinberg, and M. Avignon, Phys. Rev. Lett. {\bf 101}, 107001 (2008).

\bi{Zaz09} A. Zazunov, R. Egger, T. Jonckheere, and T. Martin, Phys. Rev. Lett. {\bf 103}, 147004 (2009).

\bi{Bru13} A. Brunetti, A. Zazunov, A. Kundu, and R. Egger, arXiv:1305.3816.

\bi{Fri52} J. Friedel, Philos. Mag. {\bf 43}, 153 (1952).

\bi{But96} M. B\"uttiker and C.A. Stafford, Phys. Rev. Lett. {\bf 76}, 495 (1996). 

\bi{Shi10} V. Shivamoggi, G. Refael, and J.E. Moore, Phys. Rev. B {\bf 82}, 041405 (2010).

\bi{Fle10} K. Flensberg, Phys. Rev. B {\bf 82}, 180516 (2010).

\end{thebibliography}
\end{document}